\documentclass[aps,prl,twocolumn,amsfonts,showpacs,superscriptaddress]{revtex4} 
\usepackage{bm}
\usepackage{epsfig,amsopn}
\usepackage{graphicx}
\usepackage{amsmath}
\usepackage{natbib}
\usepackage{color}
\usepackage{latexsym}
\usepackage{epsf}
\usepackage{amsfonts}
\usepackage{amssymb}

\renewcommand{\section}[1]{{\par\it #1.---}}
\def\beq{\begin{eqnarray}}
\def\eeq{\end{eqnarray}}

\def\nn{\nonumber}
\def\p{\partial}
\begin{document}
\title{Waiting for rare entropic fluctuations}
\author{Keiji Saito}
\affiliation{Department of Physics, Keio University, Yokohama 223-8522, Japan} 

\author{Abhishek Dhar}
\affiliation{International centre for theoretical sciences, TIFR, IISC campus, Bangalore 560012}

\date{\today}

\begin{abstract}
Non-equilibrium fluctuations of various stochastic variables, such as work and entropy production, 
have been widely discussed recently in the context of large deviations, cumulants and fluctuation relations. Typically, one looks at the distribution of these observables, at large fixed time. To characterize the precise stochastic nature of the process, we here address the distribution in the time domain.  
In particular, we focus on the first passage time distribution (FPTD) of entropy production, in several realistic models. We find that 
the fluctuation relation symmetry plays a crucial role in getting the typical asymptotic behavior. Similarities and differences to the simple 
random walk picture are discussed. 
For a driven particle in the ring geometry, the mean residence time is connected to the particle current and the steady state distribution,  and it leads to a 
fluctuation relation-like symmetry in terms of the FPTD.

\end{abstract}

\maketitle 
\section{Introduction}
The past two decades have witnessed significant development in nonequilibrium thermodynamics \cite{review1,review2,review3,review4}.
The fluctuation relations are remarkable discoveries which have quantitatively 
refined the concept of the second law \cite{evans93, kurchan98,maes99,crooks00,seifert05} as applied to small systems.
One of the central issues in nonequilibrium statistical physics has been in 
characterizing the universal nature of fluctuations of thermodynamic variables, such as heat and work that quantify entropy generated in non-equilibrium processes.  
Usually, one  measures the accumulated entropic variable, say  $X$, over  
a fixed time interval $\tau$, and its  fluctuations  are then characterized through a distribution $P(X)$. 
Defining, for example, $X$ as the stochastic total entropy, 
one can then prove the detailed and integral type of fluctuation relation for any fixed time interval $\tau$,  in various Markov processes \cite{seifert05}.
For large observation times, one finds the large deviation form $P(X) \sim e^{\tau h(X/\tau)}$ \cite{hugo}, where $h(x)$ is the large deviation function. 
The corresponding cumulant generating function (CGF) is  defined by $\mu (\xi)= \log\langle e^{\xi X } \rangle /\tau$ where $\langle ...\rangle$
is an average over the steady state, and this generates the $n$th order of cumulant $I_n$ as 
\begin{eqnarray}
I_n &=& {\partial^n  \mu (\xi ) / \partial \xi^n }\, |_{\xi =0} \,. 
\end{eqnarray}
For physical quantities related to entropy production,
it is well known that the CGF shows the fluctuation relation symmetry \cite{gc,lebo99}. 
This symmetry is not only mathematically beautiful but also physically important since it reproduces  linear response results and also gives nontrivial relationships on nonlinear responses \cite{gg96,ag07,su08}.
Large deviations and the CGF 
have been  crucial  towards constructing universal thermodynamic structure of the nonequilibrium steady state \cite{bsgjl02,bd04,d07}.

The large deviation function gives us the probability of observing  rare events 
in some fixed observation time window. An interesting and natural question to ask is as to {\emph{ how long  would one have to wait to see a rare event of a specified size?}}.  This is just the question of the first passage problem for the stochastic variable $X$. 
Although the physics of fluctuation at fixed time has been intensively studied and a lot of discoveries have been made, surprisingly,  
only a little is known on the stochastic nature of its time evolution itself.
One expects that the time-evolution of stochastic thermodynamic variables should behave like a biased random walk in some configuration space, but the details of the temporal aspects have not been investigated. 

\begin{figure}
\includegraphics[width=8.5cm]{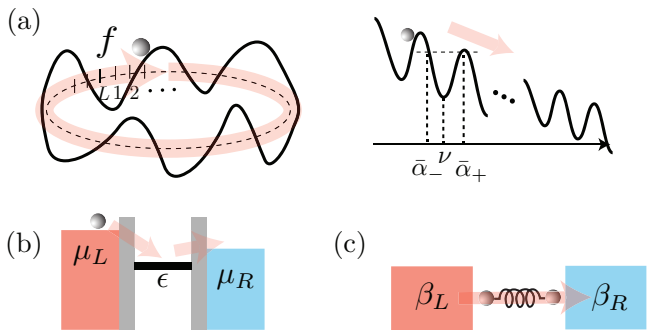} 
\caption{(color online) 
Schematic picture of setups. (a): The over-damped driven particle in the ring geometry. The right picture shows the potential landscape in an infinite line picture. (b): Classical charge transfer via a quantum-dot. (c): Heat transfer via a coupled oscillator system.}
\label{fig1}
\end{figure}

The main aim of this Letter is to investigate this aspect, which is clearly necessary for a deeper understanding of stochastic thermodynamics. 
In particular we consider the problem of the first passage time distribution (FPTD) of the desired stochastic variable, which is an experimentally measurable quantity. The FPTD here is the distribution of waiting time at which 
a stochastic variable first reaches some target value.
We consider the typical properties of the FPTD of entropy-related variables 
within the broad and well-established paradigms of stochastic thermodynamics. Three examples of nonequilibrium processes are considered: (a) an over-damped driven particle in a ring geometry, (b) classical charge transfer via a quantum-dot, and (c) heat transfer across a coupled oscillator system [see Fig.~(\ref{fig1}) for a schematic description]. 
Note that due to recent development in time-resolved measurement techniques,
there are a number of relevant experiments for these setups that look at  nonequilibrium fluctuations \cite{toyabe,bslsb07,ugmsfs10,krbmguie12,gbpc10,cilibert13}.

Using these models, we address the following questions. Is there a typical functional form for the FPTD and especially its tail ?  
How does it depend on the sign of entropy produced ? 
What are the differences as compared to the time-evolution of a simple biased random walk ?
Concerning this last question, consider the case where $X$ is the position of a biased diffusing particle on the open line. Then, defining ${\cal F}_{rw}(t,X) dt$ as the probability that the particle hits $X$ for the first time between times $t$ and $t+dt$, one easily finds \cite{redner} 
\beq
{\cal F}_{rw} ( t , X ) &=&{|X| e^{-{(X - I_1 t )^2 / 2 I_2 t }}\over \sqrt{2\pi I_2 t^3} } \to 
e^{- {I_1^2 \over 2 I_2} t - {3\over 2}\log t } \, . ~~~~~  \label{fbr}
\eeq
We will use this as a reference form, and aim to 
figure out the similarities and dissimilarities between this  simple random walk picture, and the real stochastic time-evolution of entropic variables. 
Intriguingly with use of the fluctuation relation symmetry,
one can derive the asymptotic form of the FPTD for these models,
and dissimilarities to ${\cal F}_{rw}$ can be argued. 
In addition, we derive the exact expression for mean residence time for the driven particle in 
the ring geometry which leads to fluctuation-relation-like equality in terms of the FPTD. 

\section{Driven particle in the ring geometry} 
We consider a  colloidal particle driven by a constant force and confined to move on a  ring, as depicted in Fig.~(1a).  The dynamics  is well-described  by the over-damped Langevin equations with temperature $\beta^{-1}$. The Boltzmann constant is set to unity and let us also set $ f > 0$. 
To proceed, we discretize the space into $L$ sites on the ring separated by a small spacing $a$. 
Let $P_\nu (t)$ be the probability to find the particle on the $\nu$th site at time $t$. Its evolution is given by 
\beq
{\partial P_\nu (t) \over \partial t} \!\! &=& \!\! W_{\nu,\nu-1} P_{\nu-1} (t) + W_{\nu,\nu+1} P_{\nu+1} (t) -W_{\nu,\nu} P_\nu (t)\, ,  \nonumber \\
\label{tev}
\eeq
where $W_{\nu,\alpha}$ is the transition rate matrix element which satisfies the local detailed balance condition 
$W_{\nu+1,\nu}/W_{\nu,\nu+1}=e^{-\beta (U_{\nu+1} - U_{\nu}) + \beta a f}$ and $U_\nu$ is the potential energy at the $\nu$th site. 
It is useful to introduce the winding number, ${ N}$,  which, for any given particle trajectory,  is obtained by counting the number of times the particle makes the transition from site $L$ to the first site (reverse transitions from the first site to the site $L$ are counted with a negative sign). The particle's state can be  labeled by the duplet $(\nu,{ N})$.
Suppose that in any given realization of the stochastic process, the particle 
makes the transition $(\nu,0) \to (\alpha, N)$ in  time $t$. Then the work done is $w=f ( { N} L +\alpha -\nu)$ while the heat dissipated into the bath is $q=w-U_\alpha+U_\nu$. 
Thus, at sufficiently large times, entropy production rate is proportional to the average rate of the winding number. Due to the positive force $f$, the particle on average moves in the positive direction and the average winding number rate is positive. 
However  there is a finite probability to observe the particle  moving in the opposite direction. 
The ratio of probabilities between positive and negative winding number at any finite time is quantitatively given by the fluctuation relation. We 
address the FPTD for the winding number. 
Let $T_{\alpha,\nu} ({ N} , t)$  be the transition probability from $(\nu,0)$ to $(\alpha,{ N})$ and let $F_{\alpha,\nu} ({ N} , t)$ be the probability that first passage between  $(\nu,0)$ to $(\alpha,{ N})$ occurs between time $t$ to $t+dt$. 
We note the relation for ${ N}\ne 0$ 
\beq
T_{\alpha,\nu} ({ N}, t) &=& \int_{0}^t \, du \, T_{\alpha,\alpha} (0, t-u) F_{\alpha , \nu }
 ({ N} , u)  \, . ~~ \label{ttf}
\eeq
Taking the Laplace transformation 
$T_{\alpha,\nu} ({ N},s) = \int_{0}^{\infty} dt\, e^{-s t } T_{\alpha,\nu} ({ N},t)$
and similarly for the FPTD we get 
\beq
F_{\alpha, \nu} ({ N}, t) &=& {1\over 2\pi i} \int_{c - i\infty}^{c + i \infty} ds \, e^{st}\,  
\frac{T_{\alpha,\nu} ({ N},s)}{T_{\alpha,\alpha } (0,s)}  . ~~~~\label{ftt-1} 
\eeq  
We consider the asymptotic behavior of the FPTD at sufficiently large waiting time.
To this end, one can write the time-evolution equation of the joint probability  of the variables ${\nu,  N}$. Solving this through Fourier-Laplace transformation, one can get the formal expression for the transition probability matrix \cite{suppl}
\beq
{\bm T} ({ N} ,s) &=& {1\over 2 \pi i} \oint {dz \over z^{{ N}+1} } {{\bm A}(z,s) \over \det \left[ s - 
{\bm W}_z 
\right] }  \, , \label{tns}
\eeq
where the matrix ${\bm W}_z$ is given by the matrix ${\bm W}$ replacing $(1,L)$ and $(L,1)$ elements by $z W_{1,L}$ and $z^{-1}W_{L,1}$ respectively, and ${\bm A}(z,s)$ is the co-factor matrix for the matrix of denominator. 
There are two singular values $z_{\pm} (s)$ from the denominator, which are connected to each other by the
fluctuation relation symmetry \cite{suppl}
\beq
z_+ (s) z_-(s) &=& e^{-\beta f La}  \, , \label{fr1}
\eeq 
where $\beta f L a$ is  the entropy   produced in the reservoir for a single winding around the ring.
Setting $\alpha=\nu$,  using the above  symmetry and Eq.~(\ref{ftt-1}), one can 
express the distribution  in terms of only one singular point \cite{suppl}
\beq
F_{\nu , \nu} ({ N}, t) =  {{\cal C}_{\nu} ( { N})\over 2\pi i} \int ds \, 
e^{t  \left[ s + b\, \log z_+ (s) \right] } \, , \label{asym0}
\eeq 
where $b={ N}/t$. The steadty state FPTD of winding is then given by 
${\cal F} (N ,t)=\sum_{\nu} F_{\nu , \nu} (N , t)\, p_{\nu}^{SS}$, with 
the steady state distribution $p_{\nu}^{SS}$.
A further careful examination reveals that the singular value $z_+ (s)$ is connected to the CGF, $\mu(\xi)$, for the winding number 
\beq
s-\mu (\xi )= 0~, \, {\rm where}~~ \xi = \log z_+ (s)~. \label{cond2}
\eeq
Based on these relations, a saddle point analysis leads to the following exact asymptotic expression of the FPTD
$\propto {\cal F}_{\rm asym} (t)$ for sufficiently large waiting time \cite{suppl}
\beq
{\cal F}_{\rm asym} (t) &=&
\exp[- {\Gamma} t -(3/2)\log t] \, , \nonumber \\
\Gamma &=& \sum_{n=0}^{\infty} { (- I_1 )^{n+2}  \over ( n+2 ) ! }
\left.  q_{n} (\xi ) \right|_{\xi =0}  \label{fptd}
\\
&=& {I_1^2 \over 2 I_2} + {I_3 I_1^3 \over 6 I_2^3}
+ { (3 I_3^2 - I_2 I_4 ) I_1^4 \over 24 I_2^5} 
+ \cdots \, , \nonumber 
\eeq
where the function $q_n (\xi )$ is connected to the CGF; $q_{n} (\xi )=
\left[({d^2 \mu (\xi) \over  d\xi^2})^{-1}{d \over d \xi }\right]^n ({d^2 \mu (\xi) \over  d\xi^2})^{-1}$.

Some important observations on Eq.~(\ref{fptd}) are now in order.
The asymptotic temporal decay form depends neither on the sign nor on the amplitude of the winding number, although the actual probability of negative and positive winding numbers differ by exponential  factor (in entropy produced). 
Thus, even extremely rare events follow the same asymptotic form.
In the linear response regime with small first cumulant, the asymptotic behavior is well-explained by the simple random walk picture ${\cal F}_{rw}$. In the far-from-equilibrium regime, however,  
critical deviation from this picture reveals itself in the higher order terms with nontrivial expressions.
This deviation will be significant in small systems where the degree of nonequilibrium is easily increased.
We note that the asymptotic form is given by the general form, in terms of cumulants, irrespective of detailed potential forms. This indicates that it
might be applicable to wider classes of physical situations.
As we see below, it turns out that the expression is valid for many other situations when the cumulants are calculated for appropriate physical quantities. 

\section{Two other examples}
We now show that the asymptotic form (\ref{fptd}) also appears for open nonequilibrium systems such as (b) classical charge transport via a quantum-dot and (c) heat transfer via coupled oscillators [See the Figs.~(1b,1c) for schematic pictures]. 

Case (b): Let $\mu_L$ and $\mu_R$ be respectively the chemical potential for the left and right leads and consider spin-less electrons  transmitted via a quantum-dot with an onsite energy $\epsilon$.
We measure transmitted electron at the right contact to the reservoir, and let
the accumulated electron transfer till time $t$ be ${n}$. 
Charge transfer produces Joule heating and is directly connected to entropy production rate as 
$\langle \dot{\cal S} \rangle = \beta (\mu_L - \mu_R ) \langle \dot{ n} \rangle $.

We now consider the FPTD of the accumulation of electron number, an experimentally measurable quantity. 
Let  $1$ and $2$  respectively denote 
the unoccupied and occupied states of the quantum-dot. Then the time-evolution of the two states is
described by the same type of dynamics as in Eq.~(\ref{tev}). 
The transition probability $W_{i,j}$ is composed of two contributions from the left and right reservoirs
${\bm W} = {\bm W}^L + {\bm W}^R$,
where $W_{1,2}^r=\gamma\left[1 - f_r (\epsilon ) \right]$ and $W_{2,1}^r=\gamma f_r (\epsilon )$ where $f_r$ is the Fermi distribution
of the $r$th lead ($r=L,R$). Hence these elements satisfy the detailed balance $W_{1,2}^{r}/W_{2,1}^{r} = 
e^{\beta (\epsilon - \mu_r)}$. 
The modified transition probability matrix ${\bm W}_z$ in (\ref{tns}) is given by the matrix ${\bm W}$ on replacing $W_{1,2}^{R}$ and $W_{2,1}^{R}$ in the 
off-diagonal matrix elements by 
$W_{1,2}^{R} z$ and $W_{2,1}^{R} z^{-1}$ respectively. 
The singular points in the denominator are $z_{\pm} (s)$ which are again connected by the fluctuation relation symmetry 
\beq
z_-(s) z_+ (s) &=& e^{-\beta (\mu_L - \mu_R)}  \, .
\eeq
In the present example it is easy to see that the first passage from the initial state $(i,{ n}=0)$ to any  fixed desired value of  ${n}$, also fixes the final 
configuration $j$. Using the renewal equation, we can obtain FPTD from 
$(i,0) \to (j,{ n})$, using the same argument as for the driven particle in the ring geometry, and find that the FPTD is proportional to Eq.~(\ref{fptd}) where now the  cumulants are for charge transfer and known exactly (see \cite{suppl}). 
For the case of many sites with strong onsite-Coulomb interaction, one may employ the symmetric simple exclusion process \cite{roche05}. 
 We can  demonstrate that the same expression is obtained analytically for this system of  coupled quantum-dots.

Case(c): We consider the example of the coupled oscillator system, exchanging heat with two heat reservoirs at temperatures $T_L, T_R$, whose dynamics is described by the overdamped Langevin equation
\beq
\gamma \dot{x}_1 = -k x + \eta_L (t) \, , ~~~ \gamma \dot{x}_2 = k x + \eta_R (t) \, ,
\label{viscodyn}
\eeq
where $x_{1,2}$ are the positions of the first and  second particles which are coupled via spring constant $k$, and $x=x_1 - x_2$.
The noise terms $\eta_r$ satisfy the fluctuation dissipation relations $\langle \eta_r (t) \eta_{r^\prime} (t') \rangle = 2\delta_{r, r^{\prime}} 
\gamma  \beta_{r}^{-1} \delta(t-t')$. In this case we consider the heat transfer into the right bath in time $t$ and this is given by ${ Q}=\int_0^t dt' k x(t') [ k  x (t') + \eta_R (t')]/\gamma
$ and are interested in the FPTD for transition from an initial state $(x, { Q}=0)$ to a state with ${ Q}$ amount of heat transferred. 
 As in the previous examples, we can think of our system executing biased diffusion in $(x,{ Q})$ space. However in this case fixing ${ Q}$ an the initial $x$ does not fix the final position and so an extension of the formulation is required. 
A heuristic derivation is given in \cite{suppl} but the final result for the tail of the FPTD for ${ Q}$ turns out to be the same as given by  Eq.~(\ref{fptd}) where now the  cumulants for heat transfer are known exactly from \cite{visco,kundu}.

\begin{figure}
\includegraphics[width=7.5cm]{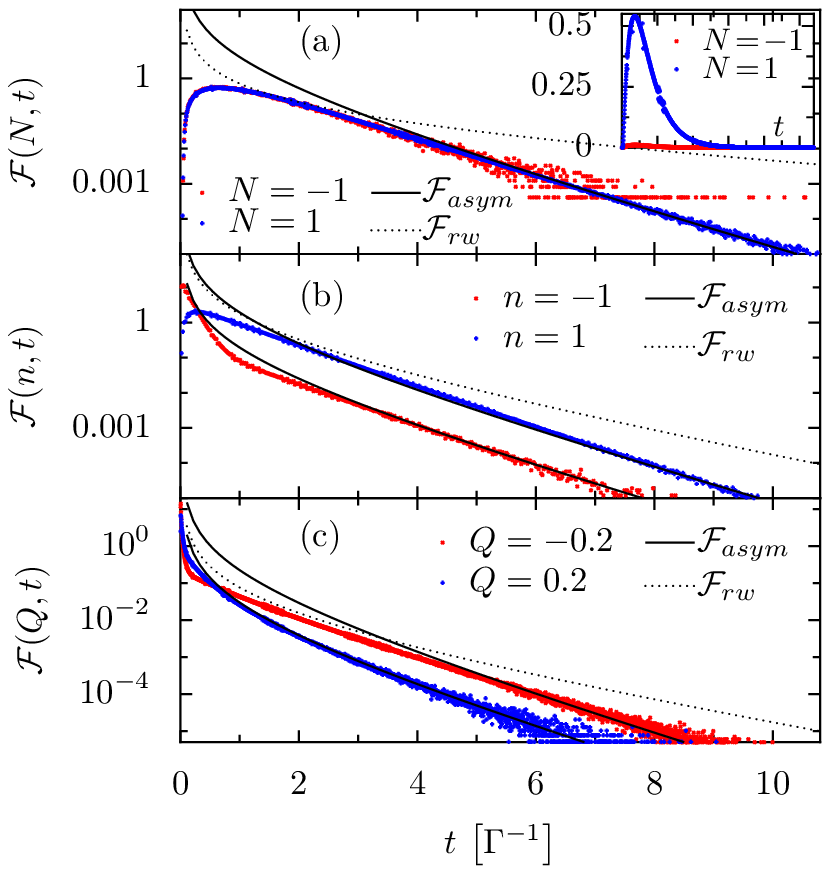} 
\caption{(color online) 
The FPTDs for the three models are shown with unit normalization. 
In the inset in (a), unnormalized data are also shown, which shows that 
the FPTD with the negative entropy production is very small. The black solid lines are fitting by 
the theory (\ref{fptd}), while the dotted line is the asymptotic curve of the random walk (\ref{fbr}).
Parameters sets are (a): $\beta=5.0$, (b): $(\mu_L , \mu_R )=(6.0,1.0)$ and (c): $(\beta_L, \beta_R)=(5.0,1.0)$. All other parameters are set to one.} 
\label{fig2}
\end{figure}
\section{Numerical demonstration of the asymptotic formula for several cases}
We numerically verify the asymptotic behavior (\ref{fptd}) for the three examples discussed and shown in  Fig.~(\ref{fig1}).  
In the case (a), we consider the dynamics in  continuous space given by
$\gamma \dot{x} = -d U(x)/ dx + f + \eta (t)$,
where we impose the periodic boundary condition with the periodic length $1$ and we employ the potential 
$U(x)=\sin (4 \pi  x)/(4\pi)$. The variable $\eta (t)$ is the Langevin noise satisfying $\langle \eta (t) \eta (t' ) \rangle =2\gamma  \beta^{-1} \delta(t-t')$.
In case (b), we numerically update the state using  a Monte-Carlo approach with the specified transition rates.  For (c) the system evolves through the Langevin dynamics in Eq.~(\ref{viscodyn}). In all cases we sample the initial state from the steady state and then measure the FPTD for  specified values of winding number [in case~(a)], the charge transfer into right reservoir [in case~(b)] and the heat transfer [in case~(c)]. 

For events with a negative entropy production there is a finite probability of the event {\emph{not occurring at all}} in a given realization. Hence for 
negative values, we plot the distribution, conditioned on the probability that 
it occurs. 
The results are shown in the Fig.~(\ref{fig2}).  In short time scale, non-universal behavior is observed. However, 
all three cases clearly show that the asymptotic behavior is well-described by 
the theory (\ref{fptd}), irrespective of the fixed values including even very rare events.
At finite times the logarithmic correction is important. 
The deviation from the simple random walk picture (which gives $\Gamma=I_1^2/(2I_2)$) is also clear.

\section{Basic equation and integral fluctuation relation in terms of first passage}
Let us consider the entropy produced in the thermal reservoir for case (a).
Let $T_{\alpha ,\nu} ({\cal S} , t)$ be the transition probability from the site $\nu$ to $\alpha$ in time 
$t$ during which the entropy of the heat reservoir increases by the amount ${\cal S}$.
The entropy is determined by the potential at the sites $\alpha$ and $\nu$ and the work done 
by external force, i.e., ${\cal S}=\beta \left[ U_{\nu} - U_{\alpha} + fa (  \alpha - \nu + L{ N} )\right]$. 
Note that fixing $\nu,{\cal S}$ does not necessarily fix $\alpha, N$. 
Let us define ${\cal F}_{\bar{\alpha},\nu} ({\cal S}, t)$ to be  the FPTD {\em only for} ${\cal S}$ [not for $(\alpha, {\cal S})$], while reaching ${\bar{\alpha}}$ from $\nu$.  For fixed ${\cal S}$, the site ${\bar{\alpha}}$ depends on $\nu$. It is uniquely determined
if  $|{\cal S}|$ is sufficiently large, while for small $|{\cal S}|$, there can be at most two choices of $\bar{\alpha}$, respectively on the two sides of $\nu$. As example, see the right figure in Fig.(\ref{fig1}a)  of a case where two $\bar{\alpha}$ (denoted by $\bar{\alpha}_{\pm}$) can be  reached for a fixed negative ${\cal S}$, starting  from the site $\nu$. Then we note the following basic equation in the Laplace representation
\beq
T_{\bar{\alpha} , \nu} ({\cal S} , s) &=&
\sum_{\bar{\alpha}^{\prime}}  T_{ \bar{\alpha} , \bar{\alpha}^{\prime}} (0 , s)\,
{\cal F}_{\bar{\alpha}^{\prime},\nu} ({\cal S}, s) \, . ~~~~~\label{hrel}
\eeq
This type of equations provides in general, a basis for considering the FPTD for entropic variables. 

We now establish several relations. The first  is an exact relation for 
the mean residence time at a given lattice point for given entropy production, 
given by
\beq 
\begin{array}{l}
\int_{0}^{\infty} dt\, {T}_{\alpha , \nu }( -{\cal S}, t) = { e^{-{\cal S}} p^{SS}_\nu \, / J }  \, , 
\\ [5pt] 
\int_{0}^{\infty} dt\, {T}_{\nu , \alpha  }( {\cal S}, t) = {p^{SS}_{\nu } \,/ J }
\, ,
\end{array} 
\label{first}
\eeq 
where $p^{SS}_{\alpha}$ is the steady state distribution at the site $\alpha$ and $J$ is the steady state particle current. In the first relation, it is  assumed that the process  $(\nu, 0) \to (\alpha , -{\cal S})$  is opposite to the direction of current, while in the second relation $(\alpha , 0)\to (\nu , {\cal S})$ is in the same direction as current. These are connected by the detailed fluctuation relation \cite{jarzynski00}. Note that the sign of ${\cal S}$ is not specified. The proof for this is presented in \cite{suppl}.
Eqs.(\ref{hrel}) and (\ref{first}) are crucial for deriving other relations as we now show. 

We now employ the usual definition 
of total entropy ${\cal S}^{\rm tot}_{ \alpha , \nu} = \ln (p^{SS}_{\nu} / p^{SS}_{\alpha }) + {\cal S}\, $.
Then for  fixed negative entropy ${\cal S} <0$, using relations (\ref{hrel}) and (\ref{first}) leads to 
the equality 
$\sum_{\bar{\alpha}} e^{ -{\cal S}^{\rm tot}_{\bar{\alpha} , \nu}} {\cal F}_{\bar{\alpha} , \nu} ({\cal S} , s=0) =1. $
Multiplying both sides of this equation by $p_{\nu}^{SS}$ and summing over $\nu$ immediately leads to the integral type of fluctuation relation 
\beq
\langle\langle e^{ - {\cal S}^{\rm tot }} \rangle\rangle_{{\cal S}} 
&=&
1 ,~~~
\label{fr-symmetry}
\eeq
where the average $\langle\langle ... \rangle \rangle_{\cal S}$ implies
taking all possible first passage paths producing the negative entropy ${\cal S}$, and that start from the steady state. Numerical demonstration is presented in \cite{suppl}.

\section{Summary}
In general, it is difficult to characterize general temporal aspects in stochastic time-evolution of thermally fluctuating objects. 
As a first step in this direction, we consider the first passage time distribution of entropy-production in several models that are relevant to recent experimental setups. We find the asymptotic behavior of Eq.~(\ref{fptd}), which seems to be the typical functional form, valid in many situations. For the paradigmatic example of a   particle driven round  a periodic potential, we find further properties, given by (\ref{first}) and (\ref{fr-symmetry}), that characterize the FPTD. It is proposed that
 Eq.(\ref{hrel}) is in general the basic equation needed while considering the FPTD for entropic variables. 

\noindent\\
{\bf Acknowledgment\hfill} \\
K.S was supported by MEXT (25103003) and JSPS
(90312983). AD thanks DST for support through the Swarnajayanti fellowship.

\clearpage

\begin{widetext}
\begin{center}
{\large \bf Supplementary Material for  \protect \\ 
``Waiting for rare entropic fluctuations'' }\\
\vspace*{0.3cm}
Keiji Saito$^{1}$ and Abhishek Dhar$^{2}$ \\
\vspace*{0.1cm}
$^{1}${\small \em Department of Physics, Keio University, Yokohama 223-8522, Japan} \\
$^{2}${\small \em International centre for theoretical sciences, TIFR, IISC campus, Bangalore 560012}
\end{center}
\end{widetext}

\section{Derivation of Eq.~(6) in the main text}
Let us define the winding number ${N}$ of a typical trajectory of the diffusing particle by counting 
the number of times it makes the transition from the site $L$ to the first site, $L\to 1$. 
The winding number decreases for the reverse transition $1 \to L$. 
Let $P_{\nu} ({N},t)$ be the joint probability that the particle is at the $\nu^{\rm th}$ site with   winding number $N$
at time $t$. We are interested in finding the probability vector 
\beq
{\bm P}({  N},t) &=& 
\left\{ 
P_1 ({  N},t) , P_2 ({  N},t) , \cdots , P_L ({  N},t)\right\} \, .
\eeq 
Then it is easy to see that this joint probability satisfies the following equation
\begin{widetext}
\beq
{\partial {\bm P} ({  N} , t) \over \partial t}  &=&
{\bm W}_{-} {\bm P} ({  N}-1 , t)+{\bm W}_{+} {\bm P} ({  N}+1 , t)
+{\bm W}_{0} {\bm P} ({  N} , t) \, . \label{suppl-joint}
\eeq
\end{widetext}
Here ${\bm W}_-$ is a $L\times L$ matrix whose only non-vanishing element is $\left[ {\bm W}_-\right]_{1,L}=W_{1,L}$,
${\bm W}_+$ is a $L\times L$ matrix whose only non-vanishing element is $\left[ {\bm W}_+\right]_{L,1}=W_{L,1}$,
and ${\bm W}_0 ={\bm W}-{\bm W}_- - {\bm W}_+$. 

We define the generating function 
\beq
{\bm P}(z,s) &=& \sum_{{  N}=-\infty}^{\infty} \int \, dt \, z^{  N} e^{-s t} {\bm P} ({  N}, t) \, .
\eeq
Then one readily find that this satisfies the equality
\beq
{\bm P}(z,s) &=& {1 \over s -{\bm W}_z } {\bm p}_0 \, 
\eeq
where ${\bm W}_z = {\bm W}_- z + {\bm W}_+ z^{-1} + {\bm W}_0$ and ${\bm p}_0$ is the initial condition ${\bm p}_0 =
{\bm P} (N=0,t=0)$. From this, one gets 
\beq
{\bm P}({  N},t) &=& {1\over 2 \pi  i } \int_{c- i\infty}^{c + i \infty} ds e^{ts} {\bm P}({  N},s) \, , \\
{\bm P}({  N},s) &=& {1\over 2 \pi i }\oint {dz \over z^{{  N}+1}} {1 \over s -{\bm W}_z } {\bm p}_0 \, .
\eeq
Hence the transition matrix ${\bm T} ({  N},s)$ is given by
\beq
{\bm T} ({  N},s)&=&{1\over 2 \pi i }\oint {dz \over z^{{  N}+1}} {1 \over s -{\bm W}_z } \, \nonumber \\
&=&{1\over 2 \pi i }\oint {dz \over z^{{  N}+1}} { {\bm A} (z,s) \over \det \left[ s -{\bm W}_z \right]} \, , 
\label{suppl-t-exp}
\eeq 
where the matrix ${\bm A} (z,s)$ is the cofactor matrix of $s -{\bm W}_z$.

\section{Derivation of the FPTD, Eq.~(10), in the main text}
\subsection{Overall structure and fluctuation relation symmetry}
In the expression (\ref{suppl-t-exp}), crucial roles are played by the singular points in the denominator. 
We first note that the determinant has the functional form
\beq
\det \left[ s -{\bm W}_z \right] 
&=& -{\prod_{k=1}^{L} W_{k+1,k}\over z} 
\left[ z - z_+ (s) \right] \left[ z- z_- (s) \right] , \nonumber \\
 \label{suppl-swz2}
\eeq
where $z_+ (s)$ and $z_- (s)$ are the singular points. By looking at the tridigonal matrix ${\bm W}_z$, one finds the relation between these singular points 
\beq
z_+ (s) z_- (s) &=& \prod_{k=1}^{L}{ W_{k,k+1} \over W_{k+1,k}} 
 =  e^{-\beta f L a} \, . \label{suppl-fl}
\eeq
Hence we can take these points as $|z_+ (s)|\ge 1$ and $|z_- (s)| \le 1$. In the limit of $s\to 0$, $z_+ (s)\to 1$.
The relation (\ref{suppl-fl}) corresponds to the fluctuation relation in terms of winding number which stands for the entropy 
$\beta f L a$  generated for every increase of the winding number.

We consider the matrix element of the transition matrix ${\bm T} ({  N},s)$. We first note that the cofactor matrix 
has dependence on either $z$ or $z^0$ or $z^{-1}$. Hence a matrix element is given by the following type of integration
\beq
{{\cal  C} \over 2 \pi i} \oint dz {z^{- n } \over \left[ z - z_+ (s) \right] \left[ z- z_- (s) \right] } \, ,
\eeq  
where $n={{  N}-1},{{  N}},{{  N}+1}$ and ${\cal C}$ is a constant dependent on the matrix element.
We here note 
\beq
\oint {dz\over 2\pi i } {z^{- n}(s)  \over \left[ z - z_+ (s) \right] \left[ z- z_- (s) \right] } 
\!\! &=& \!\!
\left\{
\begin{array}{ll}
{z_+^{-n} (s) \over 
z_- (s) - z_+ (s)  }
 \, ,  & n \ge 0 \, \\ [5pt] 
{z_-^{-n} (s) \over 
z_- (s) - z_+ (s) } \, ,  & {n} \le 0 \, 
\end{array}
\right. \, . \nonumber \\
\label{suppl-math}
\eeq 
This implies that depending the sign of the winding number ${  N}$ ($|{  N}|> 1$), the expression of numerator
takes either $z_+$ or $z_-$. However, using the symmetry (\ref{suppl-fl}) we can in unified way express 
those only with the singular point $z_+$:
\beq
T_{\alpha , \nu} ({  N}, s) &=& {\cal C}_{\alpha , \nu} (s)
{z_+^{-|n|} (s) \over 
z_- (s) - z_+ (s)  } \, , \label{suppl-tan}
\eeq
where the prefactor ${\cal C}_{\alpha , \nu}(s)$ accounts for the amplitude of transition. For instance, between positive an negative 
winding number there is exponentially large difference in the amplitude of the prefactor.

\subsection{Saddle point analysis}
Using the relation [Eq~(5) in main text] between the FPTD and the transition probability, we then get
\beq
F_{\nu , \nu } ({  N} , t) &=& {1\over 2 \pi i } \int ds e^{st} 
{T_{\nu , \nu} ({  N} , s) \over T_{\nu , \nu} (0, s)} \, \cdots {  N}\ne 0 \, . ~~
\eeq
Using Eq.(\ref{suppl-tan}), we discuss the asymptotic behavior of the FPTD. At large $t$, the function 
$F_{\nu , \nu} ({  N} , t)$ is given by
\beq
F_{\nu , \nu} ({  N} , t) &\propto & {1\over 2\pi i} \int ds e^{-t g(s,b)} \, , \label{suppl-f1}\\
g(s, b ) &=& - s + b \log z_+ (s) \, ,  \label{suppl-f2}
\eeq
where $b = {|n| / t}$. We make the saddle point analysis, where the saddle point $s^{\ast}$ satisfies
\beq
-1 + b \, { d \log z_+ (s)   \over d s } \Bigr|_{s=s^{\ast}} &=& 0 \, , \label{suppl-saddle}
\eeq
where $z_+^{\prime} (s)= d z_+ (s) / ds$. As shown in the next subsection $g^{\prime\prime}(s^{\ast})\propto 1/b^2$. 
From this, one gets
\beq
F_{\nu , \nu} ({  N} , t) &\to & e^{-t h(b=0) -(3/2)\log t} \, , \label{suppl-asym1}\\
h(b) &=& g(s^{\ast}(b) , b) = - s^{\ast} (b) + b \log z_+ (s^{\ast}(b))  ,~~~~~
\eeq
where we introduced the function $h(b)$ to emphasize that the saddle point $s^{\ast}$ is a function of $b$.
We noted in Eq.(\ref{suppl-asym1}) that asymptotic behavior implies $b\to 0$.

The function $h(b)$ is in fact precisely the large deviation function (LDF).  
To see this, we first note that the cumulant generating function (CGF) is given by the largest eigenvalue of the matrix $W_z$. Thus if $\lambda_k (z)$ is the $k^{\rm th}$ eigenvalue of ${\bm W}_z$ and  
$\lambda_{k=0}$ is, say, the largest eigenvalue then we have 
\beq
\mu (\xi ) &=& \lambda_{0} (z) \ , ~~{\rm with}~~ \xi = \log z \, .
\eeq 
The eigenvalues $\lambda_k$ are given by the roots of the determinantal eqation 
$\det\left[ \lambda - {\bm W}_z\right]=0$. This is identical to the equation for finding the roots $z_+,z_-$, namely  $\det\left[ s - {\bm W}_z\right]=0$ if we  replacing $\lambda$ by $s$.
We also note that $\lambda_0 (z) \to 0$ as $z\to 1$. 
Using this continuity in terms of the variable $s$ around $s=0$, we see that  the singular value $z_+$ is related to the CGF via the relation
\beq
 s - \mu(\xi) = 0 \, , {\rm where}~~ \xi = \log z_+ (s^{\ast}(b)) \, . \label{suppl-21}~
\eeq
Thus the function $h(b)$ is completely specified by the following equations
\begin{align}
&h(b)=b \xi(s^*) -s^*~,\label{e1}  \\
&s^*=\mu(\xi)~,\label{e2} \\
&b \frac{d \xi(s)}{d s}\big|_{s^*}-1 =0~\label{e3}.
\end{align}
From the last two equations it is easy to see that $b=d \mu/d \xi$ and hence 
it is clear that $h(b)$ is the LDF corresponding to the CGF $\mu(\xi)$.

We now express the value $\Gamma=h(b=0)$ in terms of physical quantities.
To this end, we expand the function in  a Taylor series around its maximum, $b_m$, which satisfies
\beq
d h(b)/d b \Bigr|_{b=b_m} &=& \log z_+ (s^{\ast} (b_m)) = 0 \, . \label{suppl-taylor}
\eeq
This implies $z_+(s(b_m))=1$, hence $s=0$ and therefore $h(b_m)=0$. Also $\xi=0$ and $b=d\mu/d \xi$ implies that $b_m=I_1$, the first cumulant of the winding number.  
Thus we get 
\beq
\Gamma=h (b=0) = \sum_{k=2}^{\infty} {(- I_1 )^{k} h^{(k) }(I_1) / k !} \, .
\eeq
The final task is to express the derivatives $h^{(k)}(I_1)$ in terms of cumulants of the winding number. 
To derive the expression of the second derivative $h^{(2)}(I_1)$, we start with 
the expression $h^{(2)} (b_m) = b^{-1} \, {d s^{\ast} / db} \, |_{b=b_m}$.
Using the relation $b=d\mu/d \xi$ one gets $d s^{\ast}/db |_{b=b_m}=I_1/I_2$. Hence \beq
h^{(2)} (I_1) &=& {1\over I_2} \, .
\label{suppl-h2}
\eeq
Higher order terms are systematically derived in a similar manner and we get
\begin{widetext}
\beq
h^{(2)} (b_m) &=& 1/I_2 \, \\
h^{(3)} (b_m) &=& -I_3/I_2^3 \, \\
h^{(4)} (b_m) &=& \left[ 3 I_3^2 - I_2 I_4 \right]/I_2^5 \, \\
h^{(5)} (b_m) &=& -\left[ 15 I_3^3 - 10 I_2 I_3 I_4 + I_2^2 I_5 \right]/I_2^7 \, \\
h^{(6)} (b_m) &=& \left[ 105 I_3^4 - 105 I_2 I_3^2 I_4 + 10 I_2^2 I_4^2 + 15 I_2^2 I_3 I_5 - I_2^3 I_6 \right]/I_2^9 \, \\
\vdots ~~~~~&  & \,  \nonumber  
\eeq
\end{widetext}
In conclusion we get the asymptotic behavior of the FPTD $F_{asym} (t)$
\beq
F_{asym} (t) &=& e^{-\Gamma t - (3/2)\log t} \, , \label{suppl-rr1}\\
 \Gamma &=& 
\sum_{n=0}^{\infty} { (- I_1 )^{n+2}  \over ( n+2 ) ! }
\left.  q_{n} (\xi ) \right|_{\xi =0}\, , \label{suppl-rr2} 
\eeq
where $q_{n} (\xi )=
\left[({d^2 \mu (\xi) \over  d\xi^2})^{-1}{d \over d \xi }\right]^n ({d^2 \mu (\xi) \over  d\xi^2})^{-1}$.
In the main text, we wrote the expression with up to $h^{(4)}$. 

\subsection{Logarithmic correction term: $g^{\prime\prime}(s^{\ast})\propto 1/b^2$}
We consider the equation (\ref{suppl-swz2}) for determining $z_+(s)$. We consider the structure of the quadratic equation 
\beq
z^2 + c(s) z + d &=& 0 \, , \label{suppl-qd-modela}
\eeq
where $d= \prod_{k=1}^{L}W_{k,k+1}/W_{k+1,k}=e^{-\beta f L a}$. Then the solution is 
\beq
z_+ (s) &=&  \left[- c(s) +  \sqrt{c^2 (s) - 4d }\right]/2 \, .
\eeq 
Hence the function $g(s,b)$ is given by 
\beq
g(s,b) &=& s + b \log 
\Bigl\{
 \left[- c(s) +  \sqrt{c^2 (s) - 4d }\right]/2
\Bigr\} \, .
\eeq
The first derivative is then given by 
\beq
{\partial g (s,b) \over \partial s}
&=& 1 + b 
{
{d c(s) \over ds } \left[ -1 + {c(s) \over \sqrt{c^2 (s) -4 d }}\right]
\over 2 z_+ (s)
} \, .
\eeq
Now we consider the case of $b\ll 1$. For the above to be zero, the term $\sqrt{c^2 (s) -4 d }$
 must be extremely small. 
Hence, we make the rough estimate 
\beq
\sqrt{c^2 (s^{\ast}) -4 d } \propto b \, .
\eeq
The second derivative is then estimated to be 
\beq
{\partial^2 g (s,b) \over \partial s^2} \Bigr|_{s=s^{\ast}} &\propto & {b 
\left[ c^2 (s^{\ast}) -4 d \right]^{-3/2} } \propto 1/b^2 \, .
\eeq

\section{Derivation of Eq.~(14) in the main text}
We first note that the steady state distribution and current can be exactly solved for the driven particle in the ring geometry.
\begin{widetext}
\beq
p_{\alpha}^{SS} &=& \Bigl[ 1 + {W_{\alpha, \alpha-1} \over W_{\alpha-2, \alpha-1}}
+ {W_{\alpha, \alpha-1}W_{\alpha -1, \alpha-2} \over W_{\alpha-2, \alpha-1} W_{\alpha-3, \alpha-2}}
+\cdots +
{W_{\alpha, \alpha-1}\cdots W_{\alpha -L+2, \alpha-L+1} \over W_{\alpha-2, \alpha-1} \cdots W_{\alpha-L, \alpha-L+1}} \Bigr]{\prod_{k=1}^L W_{k,k+1}\over W_{\alpha -1, \alpha}} \Bigr/{\cal Z} 
\label{suppl-steadystate}
\\
J &=& \Bigl[ \prod_{\alpha=1}^L W_{\alpha +1 , \alpha} - \prod_{\alpha=1}^L W_{\alpha , \alpha+1} \Bigr]/{\cal Z}
\label{suppl-steadycurrent} \, ,
\eeq
\end{widetext}
where we used the notation $W_{i,j}=W_{i+L,j+L}$ and ${\cal Z}$ is the normalization factor.
There are two approaches towards getting the expression of mean residence time  in terms of steady state and currents. 

In the Laplace representation for the time-domain, ${\bm P}({N},s)$, the Eq.(\ref{suppl-joint}) is reduced to
\begin{widetext}
\beq
s {\bm P}({  N},s) &=&
{\bm W}_- {\bm P}({  N}-1,s) + {\bm W}_+ {\bm P}({  N}+1,s) + {\bm W}_0 {\bm P}({  N},s) 
+ \delta_{{  N},0} \, {\bm p}_0 \, . \label{suppl-lapeq}
\eeq
\end{widetext}
For ${  N}\ne 0$, let us  of the form
\beq
{\bm P}({  N},s) &=& z^{-{  N}} {\bm V} \, ,
\eeq
where ${\bm V}$ is a constant vector. Plugging this into (\ref{suppl-lapeq}) for ${  N}\ne 0$
gives the following equation for determining $z$ and ${\bm V}$
\beq
\left[ s - {\bm W}_z \right] {\bm V} &=& 0 \, .
\eeq
A careful look at these equations reveals that there are two sets of solutions to these equations.
To see this, we write the above equation in the following form:
\beq
\left( 
\begin{array}{cc}
s +  \left[ W_{L,1} + W_{2,1} \right]  &  - {\bm Z}_+  \\
- {\bm Z}_- & {\bm U} 
\end{array}
\right)
\left( 
\begin{array}{c}
1 \\
{\bm V}^{\prime}
\end{array}
\right) &=& 0 \, .
\eeq
where 
\beq
{\bm Z}_+ &=& (W_{1,2},0,\cdots,z W_{1,L}) \, , \\
{\bm Z}_- &=& (W_{2,1},0,\cdots,z^{-1} W_{L,1})^T \, , \\
{\bm U} &=& s{\bm I}_{L-1} - {\bm W}^{2L} \, , \\
{\bm V}^{\prime} &=& (V_2, V_3, \cdots , V_L )^T \, .
\eeq
Here the matrix ${\bm W}^{2L}$ denotes $(L-1)\times (L-1)$ sub-matrix of ${\bm W}$ excluding the first row and column,
while ${\bm I}_{L-1}$ is unit matrix of dimension $(L-1)$. We have set the first element ${V}_1$ to one. Then
we get the following equations for ${\bm V}^{\prime}$ and $z$
\beq
s + W_{L,1} + W_{2,1} - W_{1,2} V_2 - z W_{1,L} V_{L} &=& 0 \, , \\
{\bm V}^{\prime} &=& {\bm U}^{-1} {\bm Z}_- \, . ~~~~~~
\eeq
The second equation leads to the relation $V_{\alpha +1}=\left[ 
U^{-1}_{\alpha 1}W_{2,1}+ U^{-1}_{\alpha L-1} z^{-1}W_{L,1}\right]$, for $\alpha = 1,\cdots,L-1$. Since
${\bm U}$ does not depend on $z$ we see that $V_{\alpha}$ are linear functions of $z^{-1}$. Hence putting back 
$V_2, V_L$ into the first equation above, we get a quadratic equation for $z$. For the two solutions we get two
corresponding explicit forms for the vectors ${\bm V}$. We denote the two solutions by 
$\{ z_+ (s) , {\bm V}^+ (s)\}$ and $\{ z_- (s) , {\bm V}^- (s)\}$. From the equation for $z$, we see the 
fluctuation relation symmetry $z_+ (s) z_-(s) = \prod_{k=1}^{L}W_{k,k+1}/W_{k+1,k}=e^{-\beta f L a} \,(<1)$.

Let us now look for a solution corresponding to the initial condition that the point starts from $\nu$ with 
${  N}=0$. A possible solution of Eq.(\ref{suppl-lapeq}) is
\beq
{\bm P}({  N},s) =
\left\{
\begin{array}{ll}
A_+ z_+^{-{  N}} {\bm V}^+ & {\rm for} ~~ {  N}>0 \, \\
A_- z_-^{-{  N}} {\bm V}^- & {\rm for} ~~ {  N}<0 \, \\
{\bm V}^0 & {\rm for} ~~ {  N}=0  \, .
\end{array}
\right.
\eeq
The unknown constants $A_+, A_-$ and ${\bm V}^0$ can fixed by requiring that our above solution satisfies 
Eq.(\ref{suppl-lapeq}) at the sites corresponding to $\nu$ and its two nearest neighbors. Clearly then
the vector ${\bm V}^0$ must have the following structure
\begin{widetext}
\beq
{\bm V}^0 &=& \left( A_- V_1^- , A_- V_2^-, \cdots , A_- V_{\nu -1}^- , 
A_0 , A_+ V_{\nu +1}^+ , \cdots , A_+ V_N^+ \right) \, .
\eeq
\end{widetext}
There are three constants $(A_- , A_+ , A_0)$ to be determined and these will follow by writing the three
special equations at the site $\nu$ and its neighbors. Let us assume, for the moment, that none of these three sites is a boundary site on the cell (i.e., $\nu-1 >1 , \nu+1 < L$). Then we get the following equations by looking at the block of ${  N}=0$
\begin{widetext}
\beq
\left\{ \left[ s + W_{\nu-2, \nu-1} + W_{\nu , \nu-1}\right]V_{\nu-1}^- - W_{\nu-1, \nu-2} V_{\nu-2}^-
\right\} A_- &=& W_{\nu-1,\nu}A_0 \, , \\
 \left[ s + W_{\nu-1, \nu} + W_{\nu +1 , \nu}\right]A_0 
- W_{\nu, \nu-1} V_{\nu-1}^-A_- - W_{\nu, \nu+1} V_{\nu+1}^+A_+
 &=& 1 \,  , \\
\left\{ \left[ s + W_{\nu, \nu+1} + W_{\nu+2 , \nu+1}\right]V_{\nu+1}^+
- W_{\nu+1, \nu+2} V_{\nu+2}^+
\right\} A_+ &=& W_{\nu+1,\nu}A_0 \, ,
\eeq
\end{widetext}
Using the equation satisfied by ${\bm V}^{\pm}$ 
which is given from the block of ${  N}\ne 0$,
we find that the first and third equations yield
\beq
A_- = A_0/V_{\nu}^- \, , && A_+ = A_0/V_{\nu}^+ \, . \label{suppl-apm}
\eeq
Plugging these into the middle equation, one gets $A_0$:
\beq
 A_0
&=& \Bigl[s +W_{\nu-1 ,\nu} +W_{\nu+1,\nu}  \nonumber \\
&&~~~~~~~~~ -W_{\nu, \nu-1} {V_{\nu-1}^- \over V_{\nu}^- }
-W_{\nu, \nu+1} {V_{\nu+1}^+ \over V_{\nu}^+
} \Bigr]^{-1} \nonumber \\
&=&  
W_{\nu, \nu-1}^{-1} \Bigl[ 
{V_{\nu+1}^- \over V_{\nu}^- }
+ {V_{\nu+1}^+ \over V_{\nu}^+
} \Bigr]^{-1} \, . \label{suppl-a0}
\eeq
Special case $s=0$. For this case the equation ${\bm W}_z {\bm V}=0$ has two solutions. One cleary 
is for $z_+ =1$ and this is the steady state solution so we can choose 
\beq
{\bm V}^+ ={\bm p}^{SS} \, .
\eeq 
The other solution for $z_- = \prod_{k=1}^{L}W_{k,k+1}/W_{k+1,k}$ is given by
\beq
{\bm V}^- \!&=& \! \left( 1, {W_{2,1}\over W_{1,2}} , {W_{2,1} W_{3,2}\over W_{1,2} W_{2,3}} ,
\cdots , {W_{2,1} \cdots W_{L,L-1}\over W_{1,2} \cdots W_{L-1,L}} 
 \right)^T \!\!\! , ~~~~~
\eeq
as can be easily verified. From (\ref{suppl-a0}) we then get 
\beq
A_{0} &=& W_{\nu, \nu +1}^{-1} \Bigl[ 
{W_{\nu+1 , \nu} \over W_{\nu , \nu + 1} } 
- 
{V_{\nu+1}^+ \over V_{\nu }^+ } 
\Bigr]^{-1}
\eeq 
Using the fact that $J=\left[ W_{\nu +1 , \nu} V_{\nu}^+ - W_{\nu,\nu+1}V_{\nu +1 }^+ \right]$ we then get 
\beq
A_0 &=& V_{\nu}^+ / J \, .
\eeq
From Eq.(\ref{suppl-apm}) we get 
\beq
\begin{array}{ll}
A_-  = {V_{\nu}^+ / J V_{\nu}^-} \, , &A_+ = {1 /  J } \, ,
\end{array}
\eeq
From this we finally obtain, for the case $s=0$, the following transition matrices for any states $\alpha, \nu$
where $\nu$ is the one ``down the hill" (i.e. current is in the direction $\alpha\to\nu$).
\beq
T_{\alpha, \alpha} ({  N}=0 , s=0) &=& {p_{\alpha}^{SS} \over J} \, , \label{suppl-talal}\\
T_{\alpha, \nu} ({  N}=0 , s=0) &=& {p_{\alpha}^{SS} \over J} 
\prod_{\nu}^{k=\alpha} 
\left( {W_{k-1, k} \over W_{k,k-1} } \right) \, , \label{suppl-alnu}\\
T_{\nu, \alpha} ({  N}=0 , s=0) &=& {p_{\nu}^{SS} \over J} \, . \label{suppl-tnual}
\eeq

We finally explain how to obtain Eq.~(14), in the main text, using Eq.~(\ref{suppl-talal}). We note that the entropy produced in the thermal reservoir for the process $\alpha\to\nu$ is given by ${\cal S} = \beta \left[ 
U_{\alpha} - U_{\nu} + fa (\nu - \alpha + LN) \right]$. This implies that the process 
$(\alpha , N=0)\to (\alpha , N=0)$ is equivalent to the process $(\alpha , {\cal S}=0)\to (\alpha , {\cal S}=0)$. Hence
Eq.(\ref{suppl-talal}) is equivalent to
\beq
T_{\alpha, \alpha} ({\cal S}=0 , s=0) = \frac{p_{\alpha}^{SS}}{J} ~.\label{suppl-ent1} 
\eeq
Now  consider the process $(\alpha , {\cal S}=0)\to (\nu , {\cal S})$ whose direction is the same as 
the average current. In this case we note $F_{\nu , \alpha}({\cal S} , s=0)=1$ which means that the process will occur with probaility one. This gives 
\beq
T_{\nu, \alpha} ({\cal S} , s=0) &=& F_{\nu, \alpha} ({\cal S} , s=0) \, T_{\nu, \nu} ({\cal S}=0 , s=0) \nonumber \\
&=&{p_{\nu}^{SS} \over  J} \, . \label{suppl-ent2}  
\eeq
In the backward process $(\nu , {\cal S}=0)\to (\alpha , -{\cal S})$ which is opposite to the direction of average current, the detailed fluctuation relation immediately gives
\beq
T_{\alpha, \nu} (-{\cal S} , s=0) &=& e^{- {\cal S}} {p_{\nu}^{SS} \over J} \, . \label{suppl-ent3}  
\eeq
These give Eq.~(14) in the main text.

\section{The FPTD of charge transfer via a quantum-dot}
The dynamics of the charge transfer in classical transport is described by
\beq
{\partial {\bm P} (t) / \partial t} 
&=& {\bm W} {\bm P} (t) \, , \\
{\bm W} &=& \sum_{r = L,R} {\bm W}^{r} \, ,
\eeq
where both ${\bm {W}}$ and ${\bm W}^r$ are $2\times 2$ matrices. $P_1 (t)$ and $P_2 (t)$ are respectively stand for the probability for unoccupied and occupied state inside the quantum-dot. Standard setup takes the transition matrix element as
$W_{1,2}^r=\bar{\gamma} \left[ 1 - f_{r} (\epsilon)\right]$ and $W_{2,1}^r=\bar{\gamma} f_{r} (\epsilon)$ where $f_{r} (\epsilon)$ is the Fermi-distribution for the $r$th reservoir and $\bar{\gamma}$ is a hopping rate. Hence it satisfies the detailed balance 
$W_{1,2}^r/ W_{2,1}^r = e^{\beta (\epsilon - \mu_r )}$. Without loss of generality, we can impose $\mu_L > \mu_R$.

Let $P_i ({Q} , t)$ be the joint probability for the $i\,(=1,2)$ state (=unoccupied or occupied state) 
and transmitted charge ${Q}$ measured at the right reservoirs till time $t$. The dynamics is given by
\begin{widetext}
\beq
\partial_t P_1 ({Q} , t) &=&
 W_{1,1} P_1({Q} , t) + W_{1,2}^L  P_2 ({Q},t ) + W_{1,2}^R  P_2 ({Q} -1 ,t )  \, , \\
\partial_t P_2 ({Q} , t) &=& W_{2,2} P_2({Q} , t) 
+
 W_{2,1}^L  P_1 ({Q},t ) + W_{2,1}^R  P_1 ({Q} +1 ,t )  \, .
\eeq
\end{widetext}
We define the generating function 
\beq
{\bm P} (z , s) &=& \sum_{{Q}=-\infty}^{\infty} \int_{0}^{ \infty} d t \,z^{Q} e^{-st} \, {\bm P}({Q} , t) \, , 
\eeq
From the dynamics for the joint probabilities, this is formally given by the expression
\beq
{\bm P}(z , s) &=& {1 \over s - {\bm W}_{z} } {\bm p}_0 \, ,
\eeq
where ${\bm W}_{z}$ is given by
\beq
{\bm W}_{z} &=& 
\left( 
\begin{array}{cc}
W_{11}  &  W_{1,2}^L + W_{1,2}^R \, z \\
W_{2,1}^L + W_{2,1}^R \,z^{-1}  & W_{22}
\end{array}
\right) \, . ~~~~~
\eeq
The transition matrix is hence given by
\beq
{\bm T} ({Q} , s) &=& 
{1\over 2 \pi i}\oint {d z \over z^{{Q}+1}}
 {1 \over s - {\bm W}_{z} } \, , \\
&=&{1\over 2 \pi i}\oint {d z \over z^{{Q}+1}}
 {{\bm A} (z , s) \over \det \left[s - {\bm W}_{z} \right] } \, .
\eeq
Now one can see the same structure to the case of ring geometry. 
From this, we find the singularities $z_{\pm} (s)$ by solving the equation $\det \left[s - {\bm W}_{z} \right]=0$, which 
are connected by the fluctuation relation symmetry 
\beq
z_{-}(s)  z_+(s) &=& {W_{1,2}^L W_{2,1}^R \over W_{1,2}^R W_{2,1}^L}=
e^{-\beta (\mu_L - \mu_R )} \, . \label{suppl-frs2}
\eeq
In the same way as in the driven particle in the ring geometry, any matrix elements of the FPTD have the following dependence
\beq
F_{\alpha, \nu} ({Q} , t ) \!&=&\!{1\over 2\pi i } \int_{c+i\infty}^{c - i\infty}   ds \, e^{ st } \,
T_{\alpha , \nu} ({Q} ,s) / T_{\alpha , \alpha}(0 ,s)  \, 
\nonumber  \\
&\propto&  {1\over 2\pi i } \int ds \, e^{-t g(s, b)}  \, , \\
g(s, b) &=& -s + b \, \log z_+ (s) \, ,
\eeq
where $b=|{Q}|/t$ and ${\cal C}$ is time-independent matrix. The argument from this point follows the calculations from Eqs.(\ref{suppl-f1}) and (\ref{suppl-f2}) in the driven particle in the ring geometry. 
Hence, we can reach the same expression as in Eqs.(\ref{suppl-rr1}) and (\ref{suppl-rr2}).

\section{Heuristic derivation of the FPTD asymptotic}
Consider a general process with discrete configuration space ${\bf X}$  and let us look at the joint  distribution of ${\bf X}$ and some quantity $Q$ (like heat or charge). This distribution, $P({\bf X}, Q,t)$, will satisfy the equation of motion 
\beq
\frac{\p P({\bf X}, Q,t)}{\p t}= {\cal L}_Q P({\bf X}, Q,t)~, 
\eeq
while the generating function $Z({\bf X}, \xi)= \int dQ e^{\xi Q} P({\bf X},Q)$ will satisfy the equation
\beq
\frac{\p Z({\bf X}, \xi,t)}{\p t}= {\cal L}_\xi Z({\bf X}, \xi,t)~.
\eeq               
The large time solution of this equation with the initial condition ${\bf X}={\bf Y}, Q=0$ at $t=0$ 
is given by
\beq
T_{{\bf X, Y}}(\xi,t) \sim e^{\mu(\xi) t} \phi({\bf X},\xi)~\chi({\bf Y},\xi)~, 
\eeq
where $\mu, \phi, \chi$ are respectively the largest eigenvalue of ${\cal L}_\xi$, and
the corresponding right and left eigenvectors.
Taking a time-Laplace transformation we get
\beq
T_{{\bf X, Y}}(\xi,s) \sim  \frac{\phi({\bf X},\xi)~\chi({\bf Y},\xi)}{[s-\mu(\xi)]}~, 
\eeq
Taking an inverse Laplace transformation in the variable $Q$, we get
\beq
T_{{\bf X, Y}}(Q,s) \sim  e^{-\xi^+(s) |Q|}~ {\phi({\bf X},\xi^+(s))~\chi({\bf Y},\xi^+(s))}~,
\label{txyqs}
\eeq
where we assume, based on empirical  observations, that only  the singularity, $\xi^+$, which satisfies the following relation, contributes: 
\beq
\mu(\xi^+)-s &=& 0 \, . \label{suppl-mu-s}
\eeq
We now consider first passage only of the variable $Q$ without caring for the configuration coordinates ${\bf X}$. 
Let us define ${\cal F}_{{\bf X},{\bf Y}}(Q,t)$ as the probability that the system 
starts from ${\bf Y}$ at time $t=0$ with $Q=0$, 
first reaches $Q$ in the  time interval $(t, t+dt)$ and is at ${\bf X}$ during that time interval. Then we have, for $Q \neq 0$,
\beq
{\cal F}_{{\bf X},{\bf Y}}(Q,s) = \sum_{\bf X'} T^{-1}_{\bf X, X'}(Q=0,s) T_{\bf X', Y}(Q,s)~.  
\eeq
 Now using Eq.~(\ref{txyqs}) and assuming that the wave-functions do not contribute to the asymptotic behaviour we  get
\beq
{\cal F}_{\bf X, Y}(Q,s) \sim e^{-\xi^+(s) |Q|}~.
\eeq
Finally, transforming back to the time domain, and doing a saddle-point analysis, we get
\begin{align}
{\cal F}_{\bf X, Y}(Q,t) &\sim e^{ - g(b) t}~, \nn \\
{\rm where}~~ g(b)&={\xi^+(s^*) b  - s^* } ~,~~~b=Q/t~, 
\end{align}
and $s^*$ is determined by 
\beq
s^*=\mu(\xi),~~~\frac{d \xi^+}{ds} b- 1=0~.
\eeq
Thus we recover the required Eqs.~(\ref{e1},\ref{e2},\ref{e3})~.

We here note that in all models (a)-(c) the equation (\ref{suppl-mu-s}) yields the quadratic equation 
of the following type
\beq
z^2(s) + c(s) z(s) + d &=& 0 \, ,  \label{suppl-quadr}
\eeq
which gives two solutions $z_{\pm} (s)$ and $z(s)$ is connected to $\xi (s)$ by the relation $\xi(s)=\log z(s)$, and 
the constant term $d$ comes from the fluctuation relation symmetry (as in Eq.(\ref{suppl-qd-modela}) for the case of driven particle in the ring geometry). Then from the same argument as in Sec.II-C, 
the same logarithmic correction is obtained. 
Hence, we get asymptotic behavior (\ref{suppl-rr1}) and (\ref{suppl-rr2}).

\section{Integral fluctuation relation in terms of first passage and numerical verification}
We consider the entropy produced in the thermal reservoir ${\cal S}$ for the driven particle in the ring geometry. 
Let ${\cal F}_{\bar{\alpha} , \nu} ({\cal S} , t)$ be the FPTD only for ${\cal S}$ [not for $(\alpha, {\cal S})$], while reaching $\bar{\alpha}$ from $\nu$. Depending on $(\nu , {\cal S})$, there are two possible situations; in the first case $\bar{\alpha}$ is unique, and in the second case there are two values $\bar{\alpha}$, as depicted 
in the right figure in Fig.~$(1a)$ in the main text. In both these cases we note the relation
\beq
T_{\bar{\alpha} , \nu} ({\cal S} , s) &=&
\sum_{\bar{\alpha}^{\prime}}  T_{ \bar{\alpha} , \bar{\alpha}^{\prime}} (0 , s)\,
{\cal F}_{\bar{\alpha}^{\prime},\nu} ({\cal S}, s) \, . ~~~~~\label{supplhrel}
\eeq
Then we consider $\sum_{\bar{\alpha}} e^{- {\cal S}_{\bar{\alpha},\nu}^{\rm tot}} {\cal F}_{\bar{\alpha},\nu} ({\cal S},s=0)$ for {\em negative ${\cal S}$}.
We first consider the case where $\bar{\alpha}$ is unique. Note that in this case $(\nu, {\cal S}=0)\to (\bar{\alpha} , {\cal S})$ is opposite to the direction of the average current. Then 
\beq
e^{- {\cal S}_{\bar{\alpha},\nu}^{\rm tot}} {\cal F}_{\bar{\alpha},\nu} ({\cal S} , s=0)
&=& e^{- {\cal S}} {p_{\bar{\alpha}}^{SS} \over p_{\nu}^{SS} } \, 
{ T_{\bar{\alpha},\nu} ({\cal S} , s=0) \over T_{\bar{\alpha},\bar{\alpha}} (0 , s=0)} \,  \nonumber \\
&=& e^{- {\cal S}} {p_{\bar{\alpha}}^{SS} \over p_{\nu}^{SS} } \, 
{  e^{\cal S} \, p_{\nu}^{SS}  \over p_{\bar{\alpha}}^{SS} } = 1 \, ,
\eeq
where we used Eqs.~(\ref{suppl-ent1})-(\ref{suppl-ent3}) to get the final result.

We next consider the case where $\bar{\alpha}$ has two choices. We call $\bar{\alpha}_{\pm}$ these two points,  
located respectively on the positive and negative side of $\nu$ [See the right figure in Fig.~(1a) in the main text]. Using Eq.~(\ref{supplhrel}), we have the following relation
\beq
\left(
\begin{array}{l}
T_{\bar{\alpha}_+ , \nu} ({\cal S}, 0) \\
T_{\bar{\alpha}_- , \nu} ({\cal S}, 0)
\end{array}
\right)
&=&
{\bm M}
\left(
\begin{array}{l}
{\cal F}_{\bar{\alpha}_+ , \nu} ({\cal S}, 0) \\
{\cal F}_{\bar{\alpha}_- , \nu} ({\cal S}, 0)
\end{array}
\right) \, ,
\eeq
where the matrix ${\bm M}$ is given by
\beq
 {\bm M} &=&
\left(
\begin{array}{ll}
T_{\bar{\alpha}_+ , \bar{\alpha}_+ } (0, 0)  & T_{\bar{\alpha}_+ , \bar{\alpha}_- } (0, 0) \\
T_{\bar{\alpha}_- , \bar{\alpha}_+ } (0, 0) & T_{\bar{\alpha}_- , \bar{\alpha}_- } (0, 0) 
\end{array}
\right)
\eeq
We note the following expressions in terms of the steady state distribution and steady state current 
\beq
\left(
\begin{array}{l}
T_{\bar{\alpha}_+ , \nu} ({\cal S}, 0) \\
T_{\bar{\alpha}_- , \nu} ({\cal S}, 0)
\end{array}
\right)
&=& 
{1\over J}
\left(
\begin{array}{l}
p_{ \bar{\alpha}_+}^{SS}    \\
p_{\nu}^{SS} e^{\cal S}     
\end{array}
\right) \, , \\
{\bm M} &=&
{1\over J}\left(
\begin{array}{ll}
p_{ \bar{\alpha}_+}^{SS}
 & p_{ \bar{\alpha}_+}^{SS} \\
p_{ \bar{\alpha}_+}^{SS}  & p_{ \bar{\alpha}_-}^{SS}
\end{array}
\right) \, .
\eeq
Using these expressions, straight forward calculation leads to
\beq
\sum_{\bar{\alpha} = \bar{\alpha}_+ , \bar{\alpha}_-} 
e^{- {\cal S}_{\bar{\alpha},\nu}^{\rm tot}} {\cal F}_{\bar{\alpha},\nu} ({\cal S},0) = 1 \, .
\eeq
Hence, in any cases, we have the identity 
\beq
\sum_{\bar{\alpha} }e^{- {\cal S}_{\bar{\alpha},\nu}^{\rm tot}} {\cal F}_{\bar{\alpha},\nu} ({\cal S},0) = 1~.
\eeq
This immediately leads to
\beq
\langle\langle e^{ - {\cal S}^{\rm tot }} \rangle\rangle_{{\cal S}} 
&=&1 ,~~~ ({\cal S} < 0) . \label{supplintft}
\eeq

We finally present the numerical demonstration of (\ref{supplintft}) as well as the relation Eq.~(\ref{suppl-ent1}). The Langevin equation was numerically solved with the same parameters set as in Fig.~(2a) in the main text.

\vspace*{1cm}
\begin{flushleft}
\includegraphics[width=7.5cm]{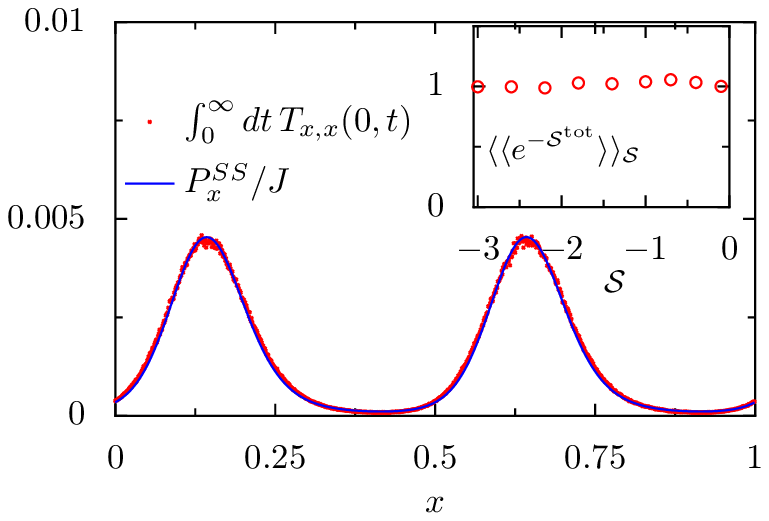} \\
{(color online) 
Numerical demonstration of Eq.(\ref{suppl-ent1}) in the main graph and fluctuation-relation-like symmetry in the inset. The Langevin equation was numerically solved with the same parameters set as in  Fig.~(2a) in the main text.}
\end{flushleft}

\end{document}